# Long range battery-less PV-powered RFID tag sensors

Sai Nithin R. Kantareddy, Ian Mathews, Rahul Bhattacharyya, Ian Marius Peters, Tonio Buonassisi, and Sanjay E. Sarma

**Abstract**— Communication range in passive Radio-Frequency Identification (RFID) front-end devices is a critical barrier in the real-world implementation of this low-cost technology. Purely passive RFID tags power up by harvesting the limited RF energy transmitted by the interrogator, and communicate by backscattering the incident signal. This mode of communication keeps manufacturing costs below a few cents per tag, but the limited power available at the tag undermines long-range deployment. In this paper, we present an approach to use Photovoltaics (PV) to augment the available energy at the tag to improve read range and sensing capabilities. We provide this extra-energy to the RFID integrated circuit (IC) using minimum additional electronics yet enabling persistent sensor-data acquisition. Current and emerging thin-film PV technologies have significant potential for being very low-cost, hence eliminating the barrier for implementation and making of PV-RFID wireless sensors. We reduce the long-range PV-RFID idea to practice by creating functional prototypes of i) a wireless building environment sensor to monitor temperature, and ii) an embedded tracker to find lost golf balls. The read range of PV-RFID is enhanced 8 times compared to conventional passive devices. In addition, the PV-RFID tags persistently transmit large volumes of sensor data (>0.14 million measurements per day) without using batteries. For communication range and energy persistence, we observe good agreement between calculated estimates and experimental results. We have also identified avenues for future research to develop low-cost PV-RFID devices for wireless sensing in the midst of the other competitive wireless technologies such as Bluetooth, Zigbee, Long Range (LoRa) backscatter etc.

✦

## 1 INTRODUCTION

The Internet of Things (IoT) market is projected to initiate growth of wireless sensor production. According to McKinsey, the overall annual value of consumer and Industrial IoT technologies will reach around $3.9 to $11.1 trillion by 2025 [1]. This will be a result of massive sensor deployment and real-time data acquisition for control, optimization and data-driven predictions in various applications. Around 67 billion IoT devices across industry segments are expected to be deployed on the network by 2025 creating a massive demand for sensor infrastructure [2]. Ultra-High Frequency (UHF) Radiofrequency Identification (RFID) technology is well positioned to provide the needed low-cost sensors [3], especially where pervasive object tracking or wireless sensing is required. Moreover, industries that have already invested in UHF RFID hardware for inventory tracking can economically leverage the existing infrastructure for other functionalities such as sensing and inventory quality monitoring.

The reader-tag communication protocol is standardized, facilitating data interoperability in sensor networks [4]. There are billions of RFID tags already in circulation that work on this globally standard Gen2 air-interface protocol in Industrial, Scientific and Medical (ISM) frequency bands. As a result, RFID technology has become popular for inexpensive, wireless object identification across a broad array of industry supply chains over the last decade [5]. RFID's ability to identify, trace and track information using easily deployable tags is now enabling applications beyond supply chain management, in new areas of sensing [6]–[8], actuation [9], and even user interaction [10]. RFID tags are used as standalone identifiers and sensors, as well as RF-front end for other off-the-shelf commercially available sensors [11], [12].

RFID tags are classified into passive, semi-passive and active tags. The integrated circuit (IC) on the passive tags is powered by harvesting the limited RF energy transmitted by the interrogator/reader, and communicate by backscattering the incident signal. Although this implementation reduces the manufacturing cost by keeping IC costs low, limited power available at the tag limits the long-range communication and power hungry sensing capabilities. In semi-passive tags, an external battery powers the IC without consuming the incident RF energy, thereby making more RF energy dispensable for backscattering over a long distance [13]. In

- *Sai Nithin R. Kantareddy is with the Department of Mechanical Engineering, Massachusetts Institute of Technology, MA- 02139. E-mail: nithin@mit.edu*
- *Ian Mathews is with the MIT PV Lab and the Department of Mechanical Engineering, Massachusetts Institute of Technology, MA- 02139. E-mail: imathews@mit.edu*
- *Rahul Bhattacharyya is with Auto-ID Labs and the Department of Mechanical Engineering, Massachusetts Institute of Technology, MA- 02139. E-mail: rahul_b@mit.edu*
- *Ian Marius Peters is with the MIT PV Lab and the Department of Mechanical Engineering, Massachusetts Institute of Technology, MA- 02139. E-mail: impeters@mit.edu*
- *Tonio Buonassisi is with the MIT PV Lab and the Department of Mechanical Engineering, Massachusetts Institute of Technology, MA- 02139. E-mail: buonassisi@mit.edu*
- *Sanjay E. Sarma is with Auto-ID Labs and the Department of Mechanical Engineering, Massachusetts Institute of Technology, MA- 02139. E-mail: sesarma@mit.edu*

active RFID, a battery-powered on-board transmitter broadcasts the RF signal without requiring any energy from the reader. However, battery-assisted wireless devices, including RFID tags, pose several design constraints such as limited lifecycle, scheduled replacement cycles, cost, size, weight, shape, lithium waste, etc. [14]–[16].

A solution to this problem is to use Photovoltaics (PV) to partially or fully reduce the dependence on batteries. There are also other non-PV ambient energy harvesting systems to power wireless-sensors [17]–[19]. However, these techniques require close proximity to the corresponding energy sources such as human body, power lines or temperature gradients. Other emerging techniques such as offloading power packets from licensed communication bands for wireless energy harvesting for IoT will require further research in high efficiency RF-DC energy converters [20]. With the widespread availability of light as an energy source both indoors and outdoors, photovoltaics are a convenient option to serve the energy needs of wireless sensors [21]–[24]. Therefore, we propose PV-RFID sensor platform that uses harvested energy from PV to powerup the IC and offers extended range compared to conventional passive RFID. Figure 1 is an illustration of flexible PV-RFID sensor communicating with RFID reader over 10-meter distance. Available literature shows the use of PV-powered RFID concept on multi-port tags (2 ICs) in [25], Intel's Wireless Identification and Sensing Platform (WISP) in [26], and Ultra-Wideband (UWB) tags in [27]. Although these studies show an increase in the read range, none of them present a persistently transmitted sensor measurement over a long time duration. Moreover, using two ICs in the multi-port tags doubles the cost compared to a single tag. PV-powered Intel's WISP is a relatively power-hungry device, which requires both RF and PV energy harvesting. On the other hand, UWB tag protocol is not standardized, and has limited commercial use. Therefore, there is a need to show PV-powered long-range wireless sensors capable of transmitting high-volume sensor measurements using an established communication protocol like UHF RFID.

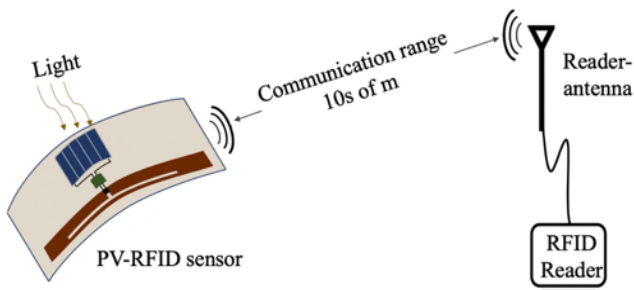

Figure 1: Schematic illustration of a PV-RFID sensor

Previously, the MIT PVLab designed and developed a solar-powered GPS tracker with cellular communication for real-world application, and demonstrated the tracker's performance on Singapore roads [28]. We also suggested how an adaptive power system for IoT sensors can reduce the battery requirement by 90% [29]. Extending such analysis to RFID would unlock ideas for blending emerging PV technologies with established low-cost wireless technology. In this paper, we show how a PV-RFID environment sensing tag that persistently measures temperature over a day is designed. We also show how a PV-RFID tag can be conformally embedded into objects such as golf balls for wireless tracking. Future data-driven IoT technologies rely on mass deployable low-cost wireless sensors that are capable of transmitting continuous sensor data. Achieving this high-volume data acquisition yet keeping the costs low is a technical challenge. We propose a simple device architecture for PV-RFID with/without intermediate energy storage to lower the costs. We reduce the idea to practice by creating functional prototypes and test the devices for read range and sensed data. In Section 2, we describe the design considerations and strategies to integrate RFID and PV into a single system. In Section 3, we discuss the results and inferences from testing the functional prototypes. Conclusions and potential future work are presented in Section 4.

## 2 DESIGN

In this section, first, components comprising a RFID system and the limitations arising from the power deficit are discussed. Next, two applications for the PV-RFID are discussed along with the design details of the prototypes. Later, how photovoltaics are suitable to fill the power gap in RFID is described.

### 2.1 RFID Tag, Range limitation and Power-gap

The tag and the reader are the two main physical components of a RFID system. The reader wirelessly interrogates the tags at speeds up to 1,000 tags/sec (theoretical rate) [30],and delivers the tag-data to the user. The tag itself is a simple impedance-matched antenna connected across an IC that processes tag's RF backscattering function by switching between the two impedance states (low and high) as a way of modulating the signal. The working principle of a UHF RFID setup is described in the additional information section.

Maximum power transmitted by the reader is constrained by the Federal Communications Commission (FCC) (or similar regional organizations) at 1 W (30 dBm), assuming an antenna with maximum of 6 dBi gain. Only a fraction of this transmitted RF power is received at the IC after path losses and polarization mismatches. According to Friis electromagnetic transmission equation, the read range is proportional to the square root of the transmission coefficient [31]. Transmission coefficient is the measure of how well the antenna's impedance matches with the IC's impedance. Transmission coefficient is affected when the tag designed for free-space environment is attached to an object with dielectric background. Lossy dielectric materials such as concrete walls, human body and water significantly affect the transmission coefficient (impedance matching) and the resulting read range. One way to solve this problem is by designing a well-matched antenna specific to the background dielectric, and an alternate approach is to use robust tag detection schemes as outlined in [32], [33].

If the RF power available at the IC is greater than the minimum power required to power-up the IC (called the IC's sensitivity), the modulated signal is backscattered, and the reader can receive the tag data. Some of the ICs, such as EM 4325 by EM Microelectronics, can function in dual modes, i.e., in both purely-passive and semi-passive modes. In semi-passive modes, this IC can take power from an external source such as PV cell. As a result, more of the RF energy is available for backscattering, which increases the read range. This extra-power from solar cell also enables continuous sensor measurements to be taken and transmitted to the reader.

Read range is compared for PV-RFID and passive RFID tags using Frii's transmission equation for free-space (described in the additional information section). Mathematically, the passive UHF tag read range performance is determined by the radiative forward link given by

$$P_{\text{IC}} = \frac{P_{TX} G_{\text{tag}} G_{\text{reader}}}{4\pi d^2} * \frac{\lambda^2}{4\pi} * \tau \ . \quad [1]$$

Therefore,

$$d = \sqrt{\frac{P_{TX} G_{\text{tag}} G_{\text{reader}} \tau}{P_{\text{IC}}}} * \frac{\lambda}{4\pi}$$

$$= constant * \sqrt{\frac{P_{TX}}{P_{\text{IC}}}}, \quad [2]$$

where $P_{\text{IC}}$: power received at the IC, $P_{TX}$: power transmitted by the reader, $d$: range or distance between the reader and the IC, $d_{\text{Passive}}$, $d_{\text{PV-RFID}}$ : range in passive and PV-RFID cases, $\lambda$: wavelength of the RF signal, $G_{\text{tag}}$, $G_{\text{reader}}$: gain of tag-antenna and reader-antennas and $\tau$: transmission coefficient.

Tag can backscatter if the power received at the IC ($P_{\text{IC}}$) is greater than the IC's sensitivity. Today, many passive RFID ICs offer sensitivities around -21 dBm. In PV-RFID, we used EM 4325 IC, which offers -31 dBm [38] in the semi-passive mode (when external power is supplied). The 10 dBm gain in the sensitivity by going from passive to semi-passive (such as PV-RFID) contributes to the extra range. Figure 2 plots the estimated range ($d$) using the above equations  with the conditions set as  $P_{\text{IC}} = -21 \ dBm$ for passive and $P_{\text{IC}} = -31 \ dBm$ for PV-RFID cases assuming $\tau = 0.8$ and overall gain $(G_{\text{tag}} G_{\text{reader}}) = 0.8$ at 915 MHz.

$$d_{\text{Passive}} = constant * \sqrt{\frac{P_{TX}}{10^{-2.1}}} \quad [3]$$

$$d_{\text{PV-RFID}} = constant * \sqrt{\frac{P_{TX}}{10^{-3.1}}} \quad [4]$$

Theoretically, ranges between 40 and 50 m are possible with PV-RFID tags, which is significantly more than the 10 m read range achievable by traditional passive tags (see Figure 2). Achieving any higher read ranges requires improvements in the reader and tag sensitivities. State-of-the-art readers' receive sensitivity is limited to -80 dBm, therefore, as the reader and tag separation is increased, after a distance, backscattered signal strength is lower than readers' sensitivity. However, the practical read range will be lower due to many factors including antenna gain, antenna-IC impedance match, reader receive sensitivity, temperature, multi-path interference, contact losses, and background dielectric impedance. With this potential increase in range, one can even monitor the footprint of an Airbus A 380's wing (~40 m) with a single reader reducing the RFID infrastructure (cables, readers, antennas, mounts, etc.) costs.

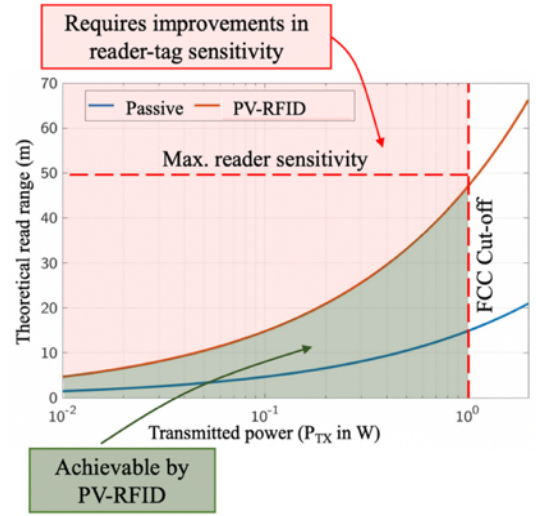

Figure 2: Comparison of read range between passive and PV-RFID tags. Note: FCC assumes an antenna with 6 dBi gain

## 2.2 Applications

We design PV-RFID devices for two cases which benefit from the extra-power at the tags: a simple attachable building environment sensor to measure window's temperature and an embedded tracker to find the lost golf balls. Figure 3 (a) is an illustration of these applications where 10-100's of battery-less sensors create a building environment sensing network and RFID augmented golf field. Every opening in the building such as windows, vents, or exhausts can be augmented with PV-RFID tags to monitor environmental parameters such as air inflow, temperature, humidity, etc. This real-time sensor data is useful to develop modern data-driven models, for example, models to run optimized and energy efficient air-conditioning systems. Figure 3 (b) shows an illustration of RFID augmented golf field, which is similar to the commercial Topgolf arenas [34].  Increase in read range due to PV-RFID decreases the number of readers and antennas required to cover the entire area of a golf field. Additionally, the extra-power can be used to power the embedded electronics to inform analytics if the player wants to know the ball spin using embedded accelerometers to analyze player's performance in the future.

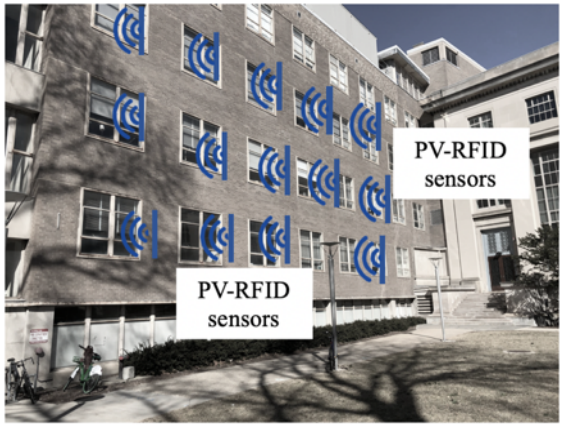

(a)

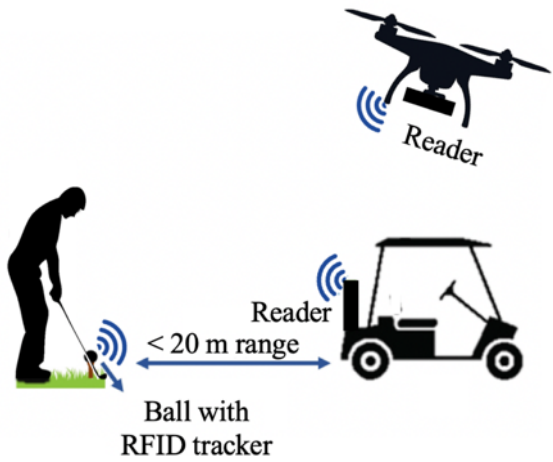

(b)

Figure 3: Illustration potential applications: building environment sensor (a); object tracker for outdoor sports (b).

### 2.2.1 Building environment sensor

The sensor consists of an antenna T-matched [35] to the EM4325 IC containing an on-board temperature sensor. An amorphous silicon (a-Si) solar cell is used to augment the power at the IC, which increases the range and capability to transmit continuous temperature measurements. A functioning prototype in use, attached to an MIT building's window, is shown in Figure 4 (a)-(b). As shown in the device's circuit schematic in Figure 4 (c), a 2.7 V Zener diode is connected across the 10 F capacitor to prevent overcharging beyond its rated capacity. Another diode is used in series with the solar cell to prevent backflow of the current when the output voltage of the solar cell is lower than the capacitor's terminal voltage. All components are assembled on an FR4 substrate with velcro backing in a form factor easy to attach to window panes, building walls or other objects. Two prototype variations are made, one with a larger capacitor size (10 F) and the other with smaller capacitor size (1 mF), to show different persistent levels in operation. Large energy backup in the former variant comes at the cost of expensive super cap and larger footprint of the product. Additionally, different types capacitors have different service times before the capacitance changes significantly from its rated value. Ceramic capacitors tend to have longer service times in the order of few year. Therefore, in real-world implementation, the designer has to carefully pick the capacitor type and size based on the cost, required output voltage, service time, required backup and product aesthetics.

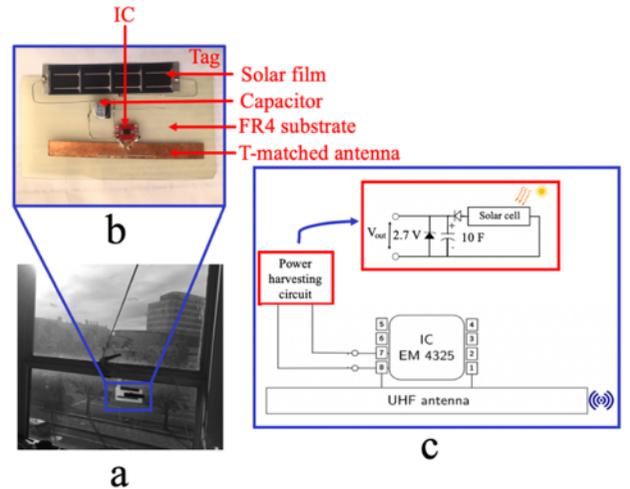

Figure 4: Prototypes of a building temperature sensor: (a) sensor attached to a window, (b) components placement, and (c) schematic of electronic circuit

### 2.2.2 Embedded tracker

PV-RFID tags are flexible and easy to embed inside objects such as a golf ball to create embeddable tags that require no frequent maintenance. We can set up 100-1000s of nodes permanently with minimum human intervention, for example, when sensors are located in hard to reach places such as the inside of sports balls, equipment, building walls, roads, etc. We demonstrate this concept of embedding PV-RFID tags, by developing an embedded tracker to find lost golf balls from a few meters distance. Embedding PV-RFID is challenging as the mass and surface topography of the golf ball have a specific function that restricts the shape, size as well as the number of embeddable electronics. In an earlier paper [36], we proposed a conformal 3D antenna design to reduce the cost of manufacturing tracker-embedded golf balls. With the conformal antenna placed between the inner core and the outer shell, these smart-golf balls can be manufactured with minimum changes to the existing manufacturing process. However, the range of the purely passive golf balls is around 1 m. With a functional prototype, we show how using PV-RFID increases the tracking range to above 10 m.

The prototype (see Figure 5 (a)-(b)) consists an embedded tracker and a flexible solar cell wrapped around the inner core to be conformal to existing design. Device's circuit schematic in Figure 5 (c) shows a solar cell directly connected to the terminals of the IC. Since an opaque golf ball does not allow the solar cell to work, we replaced the standard white outer shell with a 3D printed transparent material with transmittance

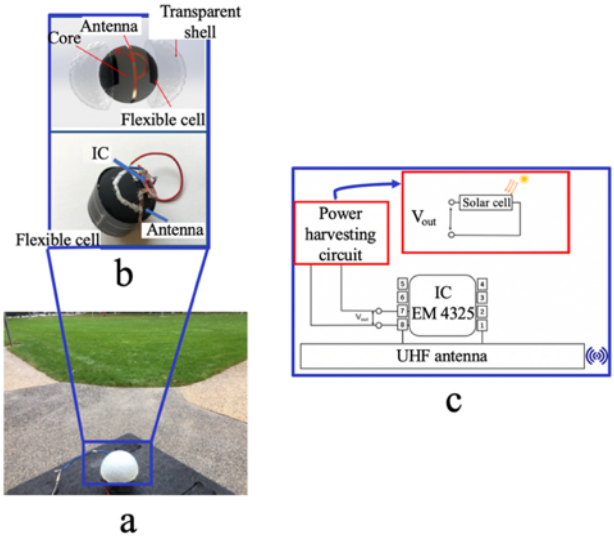

Figure 5: Prototypes of a golf ball with an embedded tracker enclosed in a transparent outer layer: (a) ball in testing, (b) internal components placement, and (c) schematic of electronic circuit

value of around 0.8. The solar cell can harvest the light reaching through the transparent layer and powerup the IC.

### 2.3 Photovoltaics to fill the power gap

Designing a PV-RFID tag involves selection of appropriate cell area, capacitor size, antenna geometry and a dielectric substrate. Solar cell area and the capacitor size affects how much energy is available at the tag and how long the tag can persist in the absence of ambient light. Antenna geometry and substrate's material should be designed so as to match the impedance of the chip for maximum backscattering of RF energy. We also looked at what percentage of the time the device can persistently operate taking into account the year-round average insolation at the location and different commercially available capacitor sizes. To achieve a 100% availability, the required minimum capacitor size is on the order of 1 – 10 F (refer additional information section). However, the large capacities affect the charging time, and with it the required cell area. Additionally, charge stored in a capacitor self-dissipates due to the leakage currents which vary from as low as 1 µA to above 50 µA. For a reasonable charging time of few minutes with the 10 F super capacitor, we select a PV module of 12 cm$^2$ area.

We also looked at how the terminal voltage in the capacitor changes over few cycles. Figure 6 shows three possible phases occurring during the operation of the PV-RFID system over a 72 hr duration: directly powered by solar cell, powered by the energy stored in the capacitor, and a blackout phase with no sufficient energy. For example, the device requires ~1.5 V and ~3.0 V of threshold voltage for read/write and EEPROM operations, respectively. We execute read/write commands to acquire the sensor-data from the PV-RFID tag, therefore, the IC consumes ~ 10 µA at 1.5 V [37]. Stored energy is limited to 2.7 V to prevent overcharging of the capacitor. This results in a sawtooth waveform with sharp charging times for the voltage across the IC as shown in the figure. At time = 0, capacitor has no charge, therefore, voltage across the IC is zero. Then the capacitor steeply charges to the maximum voltage of 2.7 V due to the energy from the solar cell. When the ambient light is available, the device is directly powered by the solar cell, as a result, the capacitor is kept fully charged and the device still operates. In the next phase, when the solar cell's output voltage drops below the capacitor's voltage, the device is then powered by the capacitor until the voltage drops below 1.5 V. In the third phase, the capacitor's terminal voltage is not sufficient to powerup the IC, therefore, the extra-range is lost until the next ambient light phase. Further analysis on the availability of the device and the effect of discharge currents can be found additional information section.

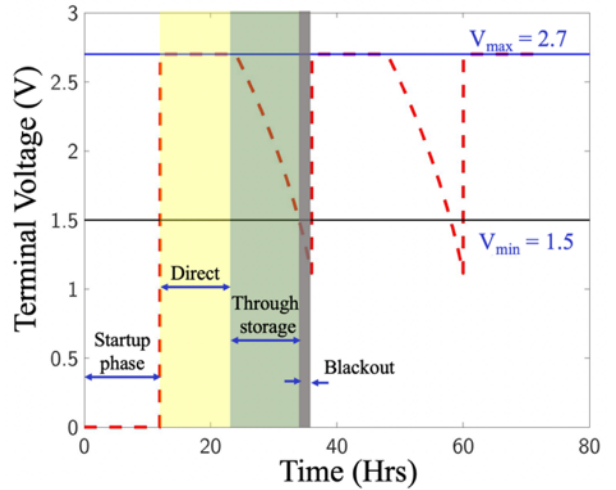

Figure 6: Showing simulated operational phases of the PV-RFID system (using a 1 F capacitor). Terminal voltage across the IC is plotted from the start of the device to next 72 hrs.

## 3 RESULTS AND DISCUSSION

In this section, we evaluate the performance of the devices, and identify few critical issues affecting the practical implementation of PV-RFID in various applications.

Read range in RFID network, or any other wireless sensor network, is a critical parameter that determines the positioning of physical gateways/readers and antennas. The infrastructure cost proportionally increases with the number of readers and antennas set up. PV-RFID increases this read range by providing external power to the IC and maximizes the backscattered signal strength. Stronger signals from the backscattering tag allows us to increase the tag-reader separation distance. This enables us to reduce the infrastructure size required to traditionally cover large spaces such as warehouses, buildings, semi-trailers, golf courses, etc. PV-RFID read range is tested using Tagformance, a standard precision RFID testing experiment by Voyantic. Figures 7 (a) and 7 (b) show the increase in read range of the building environment sensor and embedded tracker, respectively. Measured read range is consistent within +/- 2 m variation over a frequency sweep from 800 – 1000 MHz (containing standard ISM band) with 6-10 times higher than passive RFID.

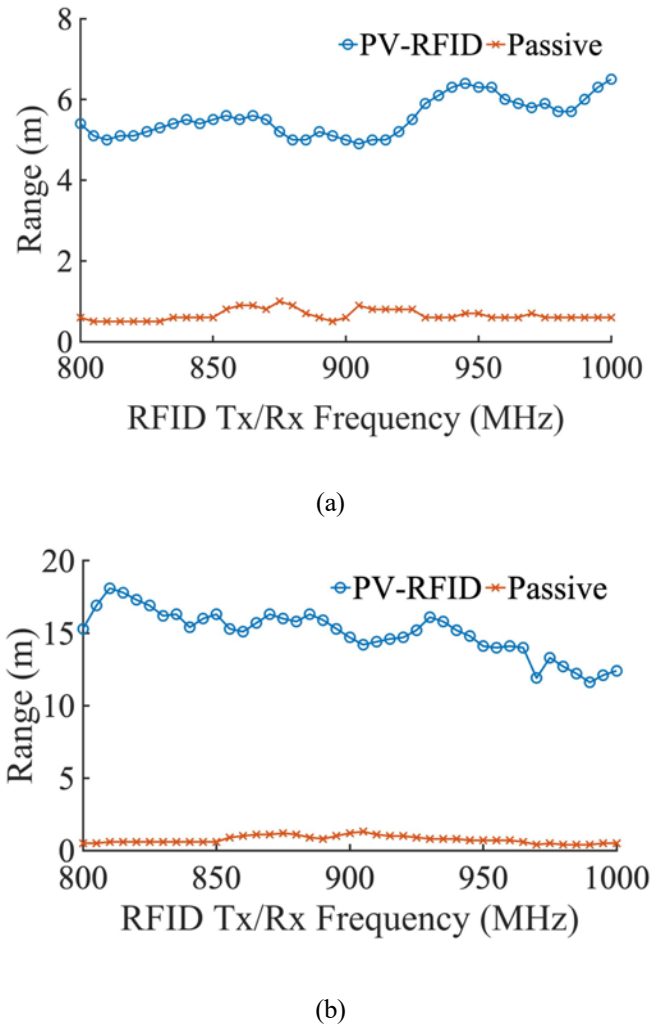

(a)

(b)

Figure 7: Increase in range of PV-RFID environment sensor (a) and embedded tracker (b)

The frequency dependent variation observed in the figures is majorly dependent on how well the tag's antenna is designed as well as on the effect of surrounding dielectric materials in the environment. This is also the reason why the same PV-RFID tag shows different boost in range in two different cases. As mentioned earlier, IC used in PV-RFID functions in two modes with two different impedances. Designing an antenna through finite element simulations to match single IC's impedance is straightforward, but performing optimization to match two different impedances needs more complex models. This complex optimization of antenna's design is not performed in this paper, resulting in variation in the read range.

The extra energy provided by PV cell to RFID also allows for continuous sensor measurements and data transmission. The environment sensor is deployed to wirelessly monitor temperature fluctuation of a window in one of the MIT's buildings (see earlier Figure 3 (a)). The sensor is interrogated by Impinj's Speedway R400 reader at an average of 4 temperature measurements per second from a distance of 2 m in an environment with 120 other random tags. The reader is placed inside the building, and the sensor is attached on the outside surface of the window using Velcro. 133,911 temperature data points are collected in a 24 hr time period using an automatic low-level reader protocol (LLRP) implementation on a modified python's sllurp library. Standard Gen 2 protocol allows for the reader to send read/write commands to the tag in the access mode. Tag's IC stores new temperature measurements in its user memory. Whenever a temperature measurement is desired, our code triggers a read command to access corresponding 4 words from the tag's user memory. If there is more than one tag in the vicinity, we select the particular tag and issue a read command. Procedure to issue read/write commands is clearly described in the LLRP protocol and even implemented in the commercial software such as Impinj's Itemtest/Multi-Reader. To emphasize the relation between the energy backup and capacitor's size, two prototypes with 10 F and 1 mF capacitors are tested, and the temperature data from these prototypes is shown in Figure 8. As expected, the data from the prototype with 10 F capacitor is continuous even during the phase with no ambient light resulting in 100%

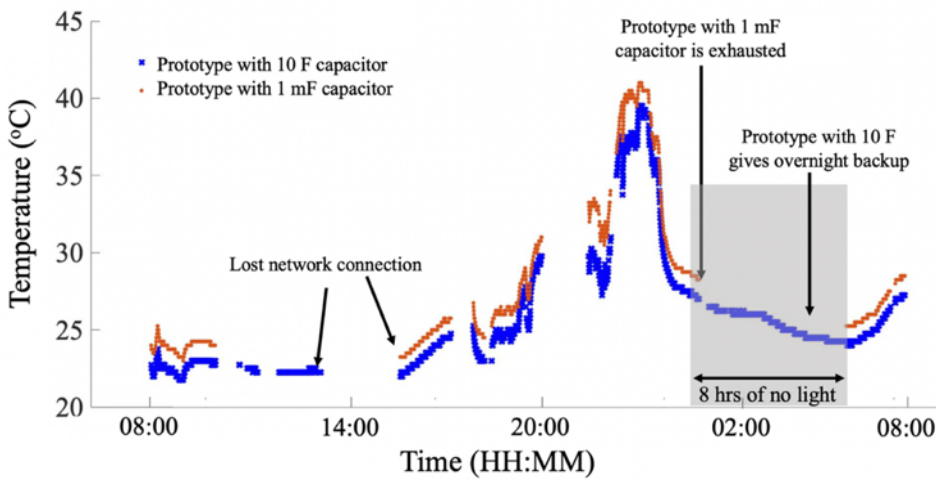

Figure 8: Temperature data of the building's window pane obtained by 133,911 measurements per sensor over 24 hr duration among other 120 random tags in the environment

persistence. The missing temperature data points in the plot correspond to the practical network connection errors, and unrelated to the energy backup. On the other hand, prototype with smaller energy backup (1 mF) stops transmitting data after 15 mins into the no-ambient-light phase. Leakage currents and the energy consumption by the IC determine how long the device can function in no-light conditions. Comparision between device's availability with leakage currents is shown in the additional information section.

In the larger capacitors, leakage currents dominate over the energy consumption by the IC. However, in the potential system-on-chip packages with tiny energy buffers, operation parameters (read/write, interrogation frequency, etc.) will significantly determine the device's availability in no-light conditions.

Replacing batteries with PV cells improves longevity of the device, reduces maintenance needed during the product life cycle, maximizes sensor-data throughput, realizes always-on tags, and reduces lithium waste. Additionally, flexible PV-RFID conforms to the shape of the objects allowing the wireless sensors with a power source to be embedded inside the objects. Currently RFID tags are mass-manufactured at around 10 cents per tag. If the current tag manufacturing is coupled with potential roll-to-roll manufacturing of emerging thin-film PV technologies, we could manufacture long read range PV-RFID tags on a single substrate at low costs. These PV-RFID tags with extended range could have potential use as access cards, markers on roads assisting self-driving cars, lightweight sensors for aerospace, deflection sensors for structural health monitoring, etc. This places PV-RFID in a competitive position with Zigbee, BLE and LoRa backscatter devices, which offer long range communication, but at higher costs, in applications requiring 10s of meters of range. However, it is important to mention that battery-powered BLE, Zigbee devices will provide higher data-rates and LoRa backscatter [38] offers even longer ranges (100s–1000s of meters range]. BLE and Zigbee offer higher data rates due to larger bandwidths, but require higher power consumption and a better energy management system than RFID. LoRa backscatter's data rate is lower than RFID, but offers longer communication ranges. It is feasible, that PV could be integrated into LoRa and other comparably lower-power communication protocols, extending their range and increasing their data-transmission frequency. PV-RFID also provides extra-power dispensable at the IC that enables new sensing modalities [7], that are not currently feasible due to the limited RF energy. Extra-power also enables battery-less high-frequency sampling required to measure vibrations in equipment maintenance. Other potential markets in the near future could be agriculture sensors, long range asset trackers, space- related sensors, etc.

## 4 CONCLUSIONS

Read range is a limitation in real-world deployment of passive RFID tags for object tracking and wireless sensing. In this paper, we examined the use of photovoltaics to increase the range by providing additional power to the tag ICs. Extra-power at the IC not only enhances the tag's ability to backscatter RF energy needed for the extra-range, but also increases the capability to measure and transmit sensor data. These flexible sensors are also suitable for embedding in 3D objects for long duration sensing where replacing batteries is not feasible. Therefore, PV-RFID enables the creation of autonomous, long range, low-cost and persistent wireless sensors.

We reduced the PV-RFID idea to practice by creating functional prototypes of a building environment sensor to measure temperature, and an embedded tracker to find lost golf balls. We tested the devices for read range and sensor data acquisition using a standard industrial RFID setup. The measured range of the PV-RFID is around 8 times higher than that of a conventional passive tag. This enables us to read PV-RFID tags spread over large areas using a smaller number of readers and antennas, decreasing infrastructure costs. The PV-RFID sensor is also shown to acquire 0.14 million measurements during a 24 hrs cycle, which shows how battery-less sensors can measure and transmit fine-grain environmental data at very low cost. For communication range and energy persistence, we observe good agreement between calculated estimates and experimental results. We have also shown that PV-RFID is flexible and easy to embed inside objects such as a golf ball to create embeddable tags that require no frequent maintenance.

In addition to increasing range and enabling continuous sensor measurements, extra energy from solar cells can enable new transduction modalities on the tag, which are not easy to execute due to limited energy at the IC. Passive tags are already mass-manufactured at low cost, therefore, by coupling with emerging thin-film PV manufacturing processes, we can potentially mass-manufacture economical long read range tag sensors.

## APPENDIX

Please see the additional information section for background material.

## ACKNOWLEDGMENTS

Authors would like to acknowledge the sources of funding for this work. S.N.R.K. has received funding from GS1 organization through the GS1-MIT AutoID labs collaboration. I.M. has received funding from the European Union's Horizon 2020 research and innovation programme under the Marie Skłodowska-Curie grant agreement No. 746516. I.M.P. was financially supported by the DOE-NSF ERF for Quantum Energy and Sustainable Solar Technologies (QESST) and by funding from Singapore's National Research Foundation through the Singapore MIT Alliance for Research and Technology's "Low energy electronic systems (LEES)" IRG.

KANTAREDDY, S.N.R. ET AL.: LONG RANGE BATTERY-LESS PV-POWERED RFID TAG SENSORS 9